\documentclass[conference]{IEEEtran}
\IEEEoverridecommandlockouts
\def\BibTeX{{\rm B\kern-.05em{\sc i\kern-.025em b}\kern-.08em
    T\kern-.1667em\lower.7ex\hbox{E}\kern-.125emX}}
\usepackage{amsmath,amssymb,mathrsfs,amsthm,bm}
\usepackage{mathtools}
\usepackage{cite}

\usepackage{graphicx,stfloats,subfigure,multicol}
\usepackage{epsfig,epsf,psfrag,amssymb,amsfonts,latexsym,graphicx,mathrsfs,subfigure}
\usepackage[table,xcdraw]{xcolor}
\usepackage{lettrine}
\usepackage{comment}
\usepackage{enumitem}
\newlist{enumsteps}{enumerate}{2}
\setlist[enumsteps,1]{label=Case \arabic*: }
\setlist[enumsteps,2]{label=Case \arabic{enumstepsi}.\arabic*: }
\usepackage{booktabs}
\usepackage[table]{xcolor}
\usepackage{multirow}
\usepackage{tablefootnote}
\usepackage{bbm}
\usepackage{orcidlink}
\usepackage{chngpage}
\usepackage{textcomp}
\usepackage{hyperref}
\usepackage{pifont}
\usepackage{url}
\usepackage[hyphenbreaks]{breakurl}
\usepackage[normalem]{ulem}
\usepackage{algpseudocode}
\newsavebox{\ieeealgbox}

\usepackage{array}

\newtheorem{theorem}{Theorem}

\newtheorem{lemma}{Lemma}
\newtheorem{definition}{Definition}

\newtheorem*{policy*}{Dynamic NEM}

 \def\old#1{}

\def\nn{\nonumber}
\def\beq{\begin{equation}}
\def\eeq{\end{equation}}
\def\bea{\begin{eqnarray}}
\def\eea{\end{eqnarray}}
\def\ba{\begin{array}}
\def\ea{\end{array}}

\def\bitem{\begin{itemize}}
\def\eitem{\end{itemize}}
\def\ben{\begin{enumerate}}
\def\een{\end{enumerate}}

\def\ie{{\it i.e.,\ \/}}

\definecolor{bgrd}{rgb}{1,1,1}
\definecolor{gray}{rgb}{0.5,0.5,0.5}
\definecolor{dkr}{rgb}{0.7,0.1,0.2}
\definecolor{dkb}{rgb}{0.1,0.1,0.8}

\def\tcr{\textcolor{red}}

\def\Cc{{\cal C}}

\def\Hc{{\cal H}}
\def\Ic{{\cal I}}

\def\Kc{{\cal K}}

\def\Mc{{\cal M}}
\def\Nc{{\cal N}}

\def\Tc{{\cal T}}

\begin{document}

\title{Resource Sharing in Energy Communities: A Cooperative Game Approach
\thanks{This work was supported in part by the National Science Foundation under Award 2218110 and the 
Power Systems and Engineering Research Center (PSERC) Research Project M-46.}
}

\author{\IEEEauthorblockN{Ahmed S. Alahmed\orcidlink{0000-0002-4715-4379} and Lang Tong\orcidlink{0000-0003-3322-2681}, \IEEEmembership{Fellow,~IEEE} (\{asa278, lt35\}@cornell.edu)\\}
\IEEEauthorblockA{\textit{School of Electrical and Computer Engineering, Cornell University}, Ithaca, USA\\}
\vspace{-0.7cm}
}

\maketitle

\begin{abstract}
We analyze the overall benefits of an energy community cooperative game under which distributed energy resources (DER) are shared behind a regulated distribution utility meter under a general net energy metering (NEM) tariff. Two community DER scheduling algorithms are examined. The first is a community with centrally controlled DER, whereas the second is decentralized letting its members schedule their own DER locally. For both communities, we prove that the cooperative game's value function is superadditive, hence the grand coalition achieves the highest welfare. We also prove the balancedness of the cooperative game under the two DER scheduling algorithms, which means that there is a welfare re-distribution scheme that de-incentivizes players from leaving the grand coalition to form smaller ones. Lastly, we present five {\em ex-post} and an {\em ex-ante} welfare re-distribution mechanisms and evaluate them in simulation, in addition to investigating the performance of various community sizes under the two DER scheduling algorithms.
\end{abstract}


\section{Introduction}\label{sec:intro}
\lettrine{T}{he} proliferating penetration levels of behind-the-meter (BTM) DER under NEM surfaced multiple issues including distribution system operator (DSO) revenue shortfalls, giving rise to successor NEM tariffs that reward DSO customers at lower rates for energy exports to the grid \cite{Alahmed&Tong:22EIRACM}. In this work, a cooperative game-theoretic formulation is developed to capture the interaction of energy prosumers in an {\em energy community} coalition and whether forming the coalition can provide its members benefits higher than those achieved under the DSO's successor NEM tariffs. Another goal is to evaluate several {\em ex-post} allocation mechanisms and compare them to the {\em ex-ante} Dynamic-NEM (D-NEM) mechanism in \cite{Alahmed&Tong:23ECjournalarXiv} that uses pricing signals to induce community members’ to achieve the community’s maximum welfare.

Cooperative game theory has been extensively used in literature to model the interaction of players and welfare/cost re-distribution in energy communities, as it allows for the analysis of coalition formation, and the allocation of benefits in a fair and efficient manner \cite{Yang&Guoqiang&Spanos:21TSG,Chakraborty&Baeyens&Khargonekar:TPS18, Han&Morstyn&McCulloch:19TPS,Fleischhacker&Corinaldesi&Lettner&Auer&Botterud:22TSG, Cui&Wang&Yan&Shi&Xiao:21TSG,Vespermann&Hamacher&Kazempour:21TPS,Chakraborty&Poolla&Varaiya:19TSG}. Various {\em ex-post}\footnote{By {\em ex-post}, we refer to the re-distribution of coalition's realized costs/welfare after its resources have been scheduled and exercised.} solution concepts have been proposed with a focus on the fairness and computational complexity of cost allocation, and whether the allocation is stabilizing, \ie no player or a subset of players would find it better off to leave the grand coalition and form smaller ones. A stabilizing nucleolus-based allocation was proposed in \cite{Yang&Guoqiang&Spanos:21TSG,Han&Morstyn&McCulloch:19TPS,Vespermann&Hamacher&Kazempour:21TPS}. Shapley and Nash bargaining-based solutions were examined in \cite{Fleischhacker&Corinaldesi&Lettner&Auer&Botterud:22TSG,Vespermann&Hamacher&Kazempour:21TPS, Chakraborty&Baeyens&Khargonekar:TPS18, Cui&Wang&Yan&Shi&Xiao:21TSG}, and a proportional rule based cost allocation was considered in \cite{Yang&Guoqiang&Spanos:21TSG,Chakraborty&Baeyens&Khargonekar:TPS18}. A mechanism that explicitly incorporates whether players are causing or mitigating cost to the coalition in allocating costs was proposed in \cite{Chakraborty&Poolla&Varaiya:19TSG}. 

Whereas some of the proposed {\em ex-post} allocation schemes were shown to belong to the cooperative game's {\em core}, including the allocation in \cite{Chakraborty&Poolla&Varaiya:19TSG}, which also satisfies the cost causation principle and has a low computation cost, the scheduling of the community DER is decoupled from the welfare/cost re-distribution problem. More specifically, the aforementioned well-established allocation schemes (e.g., Shapley, nucleolus, proportional, etc.) distribute the coalition value {\em ex-post}, which necessitates the existence of a central planner who schedules the coalition DER to maximize sharing benefits. This is unlike the D-NEM mechanism, proposed in \cite{Alahmed&Tong:23ECjournalarXiv}, which employs an {\em ex-ante} pricing rule that induces players to achieve the maximum social welfare and cost causation principle, while stabilizing the coalition.

To this end, the contributions of this work involve proposing two local energy coalitions among prosumers, where flexible demands and renewable distributed generation (DG) are optimized through two different DER scheduling algorithms: 
    \begin{enumerate}
        \item Centralized scheduling: A model that takes full control of players' DER to maximize the community's welfare.
         \item Decentralized scheduling: A model that adopts types of the DSO NEM tariff and lets players schedule their own DER.
     \end{enumerate}
   For both energy scheduling algorithms, we formulate a cooperative game and show that the value function, characterized by the coalition welfare, is {\em superadditive} and the game is {\em balanced}. For both scheduling algorithms, we evaluate five different {\em ex-post} payoff allocation mechanisms to distribute the coalition welfare and show their performance compared to the {\em ex-ante} D-NEM mechanism in \cite{Alahmed&Tong:23ECjournalarXiv}. We show that, unlike D-NEM, the {\em ex-post} allocations do not stabilize the centrally scheduled community, and require a central planner to maximize the community's social welfare.

\section{Problem Formulation}\label{sec:formulation}
\begin{figure}[htbp]
    \centering
    \includegraphics[scale=0.48]{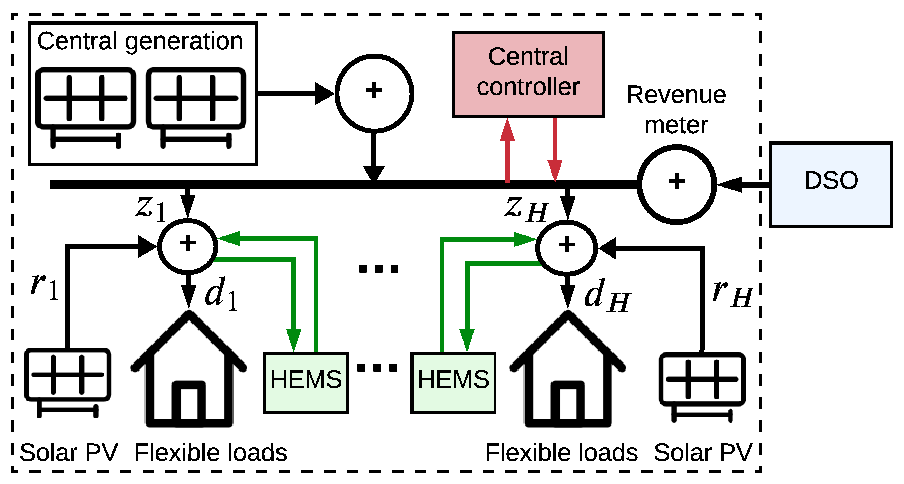}
    \setlength{\abovecaptionskip}{-0.10cm}
    \caption{Energy community coalition framework. Player's flexible consumption and renewables are $d_i, r_i \in \mathbb{R}_+$, respectively, and their net consumption is $z_i\in \mathbb{R}$. The arrows point to the positive direction of energy flow.}
    \label{fig:EnergyCommunity}
    \vspace{-0.3cm}
\end{figure}

We consider a set of $H$ electricity customers indexed by $i\in \Hc:=\{1,2,\ldots,H\}$, who may have flexible loads, BTM renewable DG, and central renewable DG. The customers face a regulated DSO that implements a NEM tariff \cite{Alahmed&Tong:22EIRACM}. This work analyzes the problem of energy coalitions in the form of {\em energy communities}, under which the members collaboratively share their DER behind the DSO's revenue meter (Fig.\ref{fig:EnergyCommunity}). As depicted in Fig.\ref{fig:EnergyCommunity}, we analyze two DER scheduling algorithms, where under the first a central controller schedules all DER by controlling the local home energy management systems (HEMS), whereas in the second, the central controller is absent, and each prosumer is independently scheduled by its local HEMS in a decentralized manner. Clearly, the privacy of members is preserved only under the latter control scheme. 

In the sequel, we mathematically model customers' DER, payment and surplus function, followed by formulating the scheduling algorithms under both scheduling algorithms.\vspace{-0.25cm}

\subsection{Member Resources, Payment, and Surplus} 

\subsubsection{Flexible Consumption, Renewable DG, and Net Consumption}
Every customer $i\in \Hc$ is assumed to have $K$ flexible loads indexed by $k \in \mathcal{K}=\{1,2,\ldots,K\}$. The {\em energy consumption bundle} of the devices is denoted by
\begin{equation*}
    \bm{d}_{i}=\left(d_{i1}, \ldots, d_{iK}\right) \in \mathcal{D}_i:=\{\boldsymbol{d}_i: \underline{\boldsymbol{d}}_i \preceq \boldsymbol{d}_i \preceq \overline{\boldsymbol{d}}_i\} \subseteq \mathbb{R}_{+}^K,
\end{equation*}
where $\underline{\bm{d}}_i$ and $\overline{\bm{d}}_i$ are the consumption bundle's lower and upper limits of customer $i$, respectively. 

The random {\em renewable DG} of every customer $i\in \Hc$, is denoted by $r_{i} \in \mathbb{R}_{+}$. Customers may own shares of the central generation, which is already accounted for in $r_i$.

The {\em net consumption} $z_i \in \mathbb{R}$ of every prosumer $i \in \Hc$ is defined as $z_{i}:= \bm{1}^\top \bm{d}_{i} - r_{i}$, where $z_{i}\geq 0$ and $z_{i}<0$ represent a {\em net-consuming} and {\em net-producing} customers, respectively. To ensure network integrity regardless of customers' net-consumption, the DSO imposes dynamic operating envelopes $\underline{z}_i \leq z_{i} \leq \overline{z}_i$, where $\underline{z}_i\leq 0$ and $\overline{z}_i\geq 0$ are the export and import envelopes, respectively \cite{Alahmed&Cavraro&Bernstein&Tong:23AllertonArXiv}. For every $i\in \Hc$, we assume $\bar{z}_i \geq \mathbf{1}^{\top} \underline{\boldsymbol{d}}_i-r_i$, and $ \underline{z}_i \leq \mathbf{1}^{\top} \overline{\boldsymbol{d}}_i-r_i$ to avoid infeasibility of the prosumer's decision.

\subsubsection{Customer Payment and Surplus}
Every $i\in \Hc$ customer is billed under the utility's NEM X regime \cite{Alahmed&Tong:22IEEETSG} based on its net consumption and the NEM X parameter $\pi=(\pi^+,\pi^-)$ as
\begin{equation}\label{eq:Pcommunity}
    P^{\pi}(z_{i})=\pi^+ [z_{i}]^+-\pi^- [z_{i}]^-,
\end{equation}
where $\pi^+, \pi^-\in \mathbb{R}_+$ are the {\em retail rate}, and {\em export rate}, respectively, and $[x]^+:=\max\{0,x\},[x]^-:=\max\{0,-x\}$ are the positive and negative parts of $x\in \mathbb{R}$. We assume $\pi^- \leq \pi^+$ as the majority of NEM successor policies consider to better align the value of excess generation with DSO deferred costs, and avoid risk-free price arbitrage \cite{Alahmed&Tong:22EIRACM, Alahmed&Tong:23ECjournalarXiv}. 

The {\em surplus} $S_{i}^{\pi}: \mathbb{R}^K \rightarrow \mathbb{R}$ of every $i\in \Hc$ is ($S_{i}^\pi(\bm{d}_{i},z_{i}) := U_{i}(\bm{d}_{i})-P^\pi(z_{i})$), where $U_{i}(\bm{d}_{i})$ is the {\em utility function} of consuming $\bm{d}_{i}$. For all $i\in \Hc$, $U_{i}(\cdot)$ is assumed to be additive (\ie $U_{i}(\bm{d}_{i})=\sum_{k\in \mathcal{K}} U_{ik}\left(d_{ik}\right)$), strictly concave, non-decreasing, and continuously differentiable with $U_{ik}(\bm{0})=0, \forall k\in \mathcal{K}$.\vspace{0cm}

\subsection{Community Energy Management Models}
Let $\Nc \subseteq \Hc$ represent an energy coalition of customers,  with cardinality $|\Nc|$, who pool their DER to improve their overall benefits and jointly share the coalition welfare (Fig.\ref{fig:EnergyCommunity}). Here, we present the two energy community control structures. 


\subsubsection{Community with Centralized Scheduling}\label{subsub:centralized}
Under this scheme, the operator of coalition $\Nc \subseteq \Hc$ maximizes community welfare by fully controlling customers' resources. Thus, the central controller solves a welfare maximization program \vspace{-0.17cm}
\begin{subequations} \label{eq:CentralOpt}
\begin{align} 
\underset{\{\bm{d}_{i}\}_{i=1}^{|\Nc|},\{z_{i}\}_{i=1}^{|\Nc|}}{\text { maximize}} &~~ W(\Nc):=\sum_{i\in \Nc} S_{i}^{\pi}(\bm{d}_{i},z_{i})\label{eq:objective} \\\text { subject to~~} & \sum_{i\in \Nc} P^\pi(z_{i}) = P^\pi(\sum_{i\in \Nc} z_{i}) \label{eq:ProfitNeut}
\\& \text{for all }~ i=1,\ldots, N\nn
\\& z_i = \bm{1}^\top \bm{d}_{i} - r_{i} \label{eq:PpiNetCons}
\\&  \underline{z}_i \leq z_{i} \leq \overline{z}_i\label{eq:Ppizit}
\\& \underline{\bm{d}}_i \preceq \bm{d}_{i} \preceq \overline{\bm{d}}_i,\label{eq:Ppidi}
\end{align} \vspace{-0.1cm}
\end{subequations}
where (\ref{eq:ProfitNeut}) is the operator's profit-neutrality constraint. Denote by $\bm{d}_{i}^{\Nc^\ast}$ and $z_{i}^{\Nc^\ast}$ the optimal consumption, and net-consumption, respectively, $\forall i \in \Nc$, that solve (\ref{eq:CentralOpt}). The optimal welfare of (\ref{eq:CentralOpt}) is therefore $ W^\ast(\Nc):= \sum_{i\in \Nc} S_{i}^{\pi}(\bm{d}^{\Nc^\ast}_{i},z^{\Nc^\ast}_{i})$.

An {\em ex-post} payoff allocation is needed to distribute the coalition welfare, as will be described in Sec.\ref{sec:Game}.

\subsubsection{Community with Decentralized Scheduling}\label{subsub:decentralized}

Under this scheme, the operator of coalition $\Nc \subseteq \Hc$ adopts the DSO's NEM X and lets its members schedule their own resources as if they were facing the DSO, \ie every $i\in \Nc$ solves\vspace{-0.2cm}
\begin{subequations} \label{eq:DecentralOpt}
\begin{align} 
\underset{\bm{d}_{i}, z_{i}}{\text { maximize}} &~~ W(i):= S_{i}^{\pi}(\bm{d}_{i},z_{i})\\ \text{subject to~~} & (\ref{eq:PpiNetCons})-(\ref{eq:Ppidi}). 
\end{align} 
\end{subequations}
\vspace{-0.10cm}
For all $i\in \Nc$, let $\bm{d}_{i}^{\dagger}$ and $z_{i}^{\dagger}$ denote, respectively, the optimal consumption, and net-consumption that solve (\ref{eq:DecentralOpt}) and result in the customer maximum welfare $W^\dagger(i):=  S_{i}^{\pi}(\bm{d}_{i}^{\dagger},z^{\dagger}_{i})$. Evidently, the total welfare of the decentralized community is suboptimal. Nonetheless, the {\em ex-ante} D-NEM in \cite{Alahmed&Tong:23ECjournalarXiv} achieves the maximum welfare, while adopting a decentralized scheduling scheme. D-NEM is briefly described in Sec.\ref{sec:Allocation}.

\section{Energy Community Cooperative Game}\label{sec:Game}
Here, we introduce the cooperative energy community game and its properties under the two DER scheduling algorithms.

\subsection{Coalitions and Value Functions}

An  {\em energy community coalition} is formed when any subset of prosumers $\Nc$ decide to cooperate by sharing their DER, and jointly face the DSO. The {\em set of all possible coalitions} is defined as the power set of $\Hc, 2^\Hc:=\{\Nc:\Nc \subseteq \Hc\}$. $\Hc $ and $\emptyset$ are the {\em grand coalition} and {\em empty coalition}, respectively.

An $|\Hc|$-player cooperative game is defined by the pair ($\Hc,\nu$), where $\nu:2^\Hc \rightarrow \mathbb{R}$ is a {\em characteristic function} that assigns {\em value} $\nu(\Nc)$ to each coalition $\Nc \subseteq \Hc$. The {\em value} function $\nu(\Nc)$ is characterized by the welfare of the energy community coalition $\Nc$
\begin{equation}\label{eq:ValueFunction}
    \nu(\Nc):= W(\Nc),
\end{equation}
with $\nu(\Nc) = 0$ if $\Nc=\emptyset$, and $\nu(\Hc)$ being the value of the {\em grand coalition}. Note that the value of a singelton coalition $\nu(i)= W(i)$ is simply the welfare of a standalone customer who autonomously faces the DSO's NEM X.

Next, we formulate the value function under both the centralized and decentralized energy management schemes. 

\subsubsection{Community with Centralized Scheduling}
Here, the $|\Nc|$ players' DER are scheduled by the community manager, which results in the maximum social welfare in (\ref{eq:CentralOpt}). From the definition in (\ref{eq:ValueFunction}), the cooperative game value function under this scheme is $ \nu^\ast(\Nc)= W^\ast(\Nc)$.

It turns out that the value function under the centralized scheme is superadditive, which means that the {\em grand coalition} achieves the highest welfare.

\begin{lemma}\label{lem:AdditiveCentral}
    The cooperative game under centralized DER control $(\Hc,\nu^\ast)$ is superadditive, i.e., $\nu^\ast(\Nc)+\nu^\ast(\Tc) \leq \nu^\ast(\Nc \cup \Tc),~~ \forall \Nc,\Tc\subseteq \Hc, \Nc \cap \Tc=\emptyset$.
\end{lemma}
\noindent {\em Proof:} See Appendix \ref{subsec:lemAdditiveCentral}.\hfill$\blacksquare$

\subsubsection{Community with Decentralized Scheduling}

Here, customers continue scheduling their DER, as if they were facing the DSO, therefore, their coalition welfare is $\sum_{i\in \Nc} S_{i}^{\pi}(\bm{d}^\dagger_{i},z^\dagger_{i})$.
Because we enforce a {\em profit-neutral} operator, we adjust this welfare by adding a term that ensures the re-distribution of any excess profit: 
\begin{align}\label{eq:Welfare1Mod}
    W^\dagger(\Nc) :=  \sum_{i\in \Nc} W^\dagger(i) + \big(\sum_{i\in \Nc} P^\pi(z_{i}^\dagger)- P^\pi(z_{\Nc}^\dagger)\big),
\end{align}
where $z_{\Nc}^\dagger:= \sum_{i\in \Nc} z_{i}^\dagger$. The second term on the right-hand side of (\ref{eq:Welfare1Mod}) is the difference between the payment the community operator collects from its members, and the payment it submits to the DSO. From (\ref{eq:ValueFunction}), the cooperative game value function is $\nu^\dagger(\Nc)= W^\dagger(\Nc)$.

Interestingly, and as formally stated in Lemma \ref{lem:AdditiveDecentral}, although the players were neither centrally scheduled, as in (\ref{eq:CentralOpt}), nor incentivized by prices to achieve the maximum welfare, as in \cite{Alahmed&Tong:23ECjournalarXiv}, the value function of their cooperation is superadditive.
\begin{lemma}\label{lem:AdditiveDecentral}
    The cooperative game under decentralized DER control $(\Hc,\nu^\dagger)$ is superadditive, i.e., $\nu^\dagger(\Nc)+\nu^\dagger(\Tc)\leq \nu^\dagger(\Nc \cup \Tc),~~ \forall \Nc,\Tc\subseteq \Hc,~ \Nc \cap \Tc=\emptyset$.
\end{lemma}
\noindent {\em Proof:} See Appendix \ref{subsec:lemAdditiveDecentral}.\hfill$\blacksquare$

\subsection{Payoff Allocation, Imputation, and Core}

We showed that both the community under both scheduling schemes has a superadditive value function, which raises the normative question of how we should assign the coalition value to each player of the cooperative game.

\begin{definition}[Payoff allocation and imputation]
    Let $\bm{\psi} \in \mathbb{R}^H$ denote the payoff allocation vector for coalition $\Nc \subseteq \Hc$, with $\psi_i$ being the payoff to prosumer $i\in \Hc$. No payoff is assigned to prosumers outside the coalition, i.e., $\psi_i=0,~ \forall i \notin \Nc$. 

    The allocation $\bm{\psi}$ is an imputation if: 1) the grand coalition welfare is wholly distributed to the members (efficiency), \ie $\sum_{i\in \Hc} \psi_i = \nu(\Hc)$, and 2) individuals are better off joining the coalition (individual rationality), \ie $\psi_i \geq \nu(i),~ \forall i\in \Hc$. Therefore, the set of all imputations is defined as
    \begin{equation}\label{eq:Imputation}
    \Ic:=\big\{ \bm{\psi}\in \mathbb{R}^H: \sum_{i\in \Hc} \psi_i = \nu(\Hc),~ \psi_i \geq \nu(i), \forall i\in \Hc \big\}.
    \end{equation}
\end{definition}
It is not hard to show that for any superadditive game, $\Ic$ is nonempty. Despite the favorable properties of a payoff allocation that is an imputation $\bm{\psi} \in \Ic$, it does not guarantee {\em group rationality}, which ensures that no subset of prosumers would be better-off if they jointly withdraw from the {\em grand coalition} to form their own coalition. Therefore, we further require the payoff allocation to be in the {\em core}, which is the fundamental solution concept for cooperative games.
\begin{definition}[Core and coalition stability]
     The core $\Cc$ for the cooperative game $(\Hc, \nu)$ is defined as the set of all payoffs that are imputations and ensures that no coalition can have a value greater than the sum of its members' payoffs
    \begin{equation}\label{eq:core}
        \Cc:= \big\{ \bm{\psi}\in \Ic: \sum_{i\in \Nc} \psi_i \geq \nu(\Nc),~ \forall \Nc\in 2^\Hc \big\}.
    \end{equation}
    The coalition is stable if $\bm{\psi}\in \Cc$. 
\end{definition}

\subsection{Balanced Games}
{\em Balanced games}, which gives a necessary and sufficient condition for the existence of a stabilizing allocation (nonempty core) in a cooperative game are defined next.

\begin{definition}[Balanced game]\label{def:BalancedGames}
    A map $\alpha:2^{\Hc} \rightarrow [0,1]$ is called a balanced map if $\sum_{\Nc \in 2^\Hc} \alpha(\Nc) \bm{1}_{\Nc}(i)=1$, where $\bm{1}_{\Nc}(i)=1$ if $i \in \Nc$ and $\bm{1}_{\Nc}(i)=0$ if $i \notin \Nc$. 
    The cooperative game ($\Hc,\nu$) is balanced if for any balanced map $\alpha$,
    $\sum_{\Nc \in 2^\Hc} \alpha(\Nc) \nu(\Nc) \leq \nu(\Hc)$.
\end{definition}

The following Theorems show that the cooperative games under both DER scheduling algorithms are {\em balanced}.
\begin{theorem}[Balancedness under central DER scheduling]\label{thm:BalancedCentral}
        The energy community cooperative game ($\Hc,\nu^\ast$) is balanced.
    \end{theorem}
    \noindent {\em Proof:} See Appendix \ref{subsec:thmAdditiveCentral}.\hfill$\blacksquare$
\begin{theorem}[Balancedness under decentral DER scheduling]\label{thm:BalancedDecentral}
       The energy community cooperative game ($\Hc,\nu^\dagger$) is balanced.
    \end{theorem}
\noindent {\em Proof:} See Appendix \ref{subsec:thmAdditiveDecentral}.\hfill$\blacksquare$\\
From the Bondareva-Shapley Theorem \cite{Shapley:63NRLQ}, the balancedness of the cooperative game under the decentralized and centralized control schemes is a necessary and sufficient condition for the non-emptiness of their core. The superadditivity and balancedness of the cooperative game imply the existence of a stable payoff allocation of the coalitional welfare that is satisfactory for every member of the grand coalition.

\subsection{Payoff Allocation Mechanisms}\label{sec:Allocation}
We present five {\em ex-post} payoff allocations and compare them to D-NEM in \cite{Alahmed&Tong:23ECjournalarXiv,Alahmed&Cavraro&Bernstein&Tong:23AllertonArXiv}, which is an {\em ex-ante} allocation mechanism. For all of the {\em ex-post} allocation schemes, we ensure that the utility of consumption $U_i(\cdot)$ is wholly assigned to each member $i$, as such value is not {\em transferable}. Hence, only the coalition payment $P^\pi(z_\Nc)$ is re-distributed. 

\subsubsection{Equal Division Allocation}\label{subsub:ED}
The equal division allocation equally splits the payment to the players \cite{HPYoung:85CostAllocationBook}. Therefore, the allocation to each player under this scheme is
\begin{equation*}\label{eq:ED}
    \psi^{(1)}_i =  U_i - P^\pi(z_\Nc)/|\Nc|,\quad \forall i \in \Nc.
\end{equation*}
\subsubsection{Egalitarian Allocation}\label{subsub:Egalitarian}
This allocation assigns to player $i$ all of his individual payment plus an equal share of any possible cost savings achieved by forming the coalition \cite{HPYoung:85CostAllocationBook}
\begin{equation*}\label{eq:Egalitarian}
    \psi^{(2)}_i =  U_i - \left(P^\pi(z_i)+\frac{P^\pi(z_\Nc)-\sum_{j=1}^N P^\pi(z_j)}{|\Nc|}\right), \forall i \in \Nc.
\end{equation*}
\subsubsection{Proportional Allocation}\label{subsub:Proportional}
The proportional allocation assigns payments to each member that is proportionate to the cost they impose on the coalition
\begin{equation*}\label{eq:Proportional}
    \psi_i^{(3)}=  U_i - \frac{W(i)}{W(\Nc) }P^\pi(z_\Nc), \quad \forall i \in \Nc,
\end{equation*}

\subsubsection{Allocation in \cite{Chakraborty&Poolla&Varaiya:19TSG}}\label{subsub:Chak}
The {\em ex-post} allocation in \cite{Chakraborty&Poolla&Varaiya:19TSG} assigns payment to players based on their net consumption and the aggregate net consumption of the community as
\begin{equation*}\label{eq:Poolla}
    \psi_i^{(4)}=   U_i - \begin{cases}
\pi^+ z_{i} & \text{ if } z_{\Nc}\geq 0 \\ 
\pi^- z_{i} & \text{ if }  z_{\Nc} < 0
\end{cases}, \quad \forall i \in \Nc.
\end{equation*}

\subsubsection{Shapley Value Allocation}\label{subsub:Shapley}
Shapley value distributes payoff among the players based on their marginal contributions to the coalition \cite{Shapley:63NRLQ} as, for every $i \in \Nc$,
\begin{align}\label{eq:Shapley}
    \psi_i^{(5)} \hspace{-0.1cm}=\hspace{-0.05cm} U_i -\hspace{-0.54cm}\sum_{\Nc \in 2^\Hc, i\in \Nc}\hspace{-0.53cm} \frac{(|\Hc|-|\Nc|) !(|\Nc|-1)!}{|\Hc| !}(P^\pi(z_\Nc)-P^\pi(z_{\Nc \setminus i})) \nn.
\end{align} 

\subsubsection{Dynamic NEM \cite{Alahmed&Tong:23ECjournalarXiv,Alahmed&Cavraro&Bernstein&Tong:23AllertonArXiv}}\label{subsub:DNEM}
Unlike the allocation mechanisms in Sec. \ref{subsub:ED}--\ref{subsub:Shapley}, which are {\em ex-post} and reliant on a centralized energy management scheme for social welfare maximization, the welfare-maximizing D-NEM employs an {\em ex-ante} and threshold-based pricing mechanism $\pi^{\mbox{\tiny DNEM}}$ that sets the community price based on the available renewables, as
\begin{equation*}\label{eq:DNEM}
    \pi^{\mbox{\tiny DNEM}}(\bm{r})=  \begin{cases}
\pi^+ & \text{ if } \sum_{i\in \Nc} r_i < f_\Nc(\pi^+)  \\ 
\pi^z(\bm{r}) & \text{ if } \sum_{i\in \Nc} r_i \in [f_\Nc(\pi^+),f_\Nc(\pi^-)] \\ 
\pi^- & \text{ if }  \sum_{i\in \Nc} r_i < f_\Nc(\pi^-).
\end{cases}
\end{equation*}
The computation of the thresholds $(f_\Nc(\pi^+),f_\Nc(\pi^-))$ and the price $\pi^z(\bm{r})\in [\pi^-,\pi^+]$ are described in \cite{Alahmed&Tong:23ECjournalarXiv} and \cite{Alahmed&Cavraro&Bernstein&Tong:23AllertonArXiv} without and with operating envelopes, respectively.  
The work in \cite{Alahmed&Tong:23ECjournalarXiv,Alahmed&Cavraro&Bernstein&Tong:23AllertonArXiv} proved that the optimal response of members under D-NEM is equivalent to the optimal response under the centralized control scheme, \ie $(\bm{d}_i^{\Nc\ast},z_i^{\Nc\ast}), \forall i\in \Nc$. Therefore, the member payoff (surplus) under D-NEM, 
\begin{equation*}\label{eq:DNEMSur}
    \psi_i^{(6)} = U_i(\bm{d}_i^{\Nc\ast}) - \pi^{\mbox{\tiny DNEM}}(\bm{r})\cdot  z_i^{\Nc\ast},
\end{equation*}
is attained rather than distributed {\em ex-post}, as members' react to the D-NEM price $\pi^{\mbox{\tiny DNEM}}(\bm{r})$.

\section{Numerical Results}\label{sec:num}
We evaluated the performance of community scheduling algorithms and the {\em ex-post} and {\em ex-ante} allocation mechanisms by constructing a hypothetical residential energy community coalition of various sizes using a one-year consumption and DER data of 20 households.\footnote{Using \href{https://www.pecanstreet.org/dataport/}{PecanStreet data} for households in Austin, TX in 2018.} The community faces a NEM policy with an hourly billing period, a ToU-based {\em retail} rate with $\pi^+_{\mbox{\tiny ON}}=\$0.40$/kWh and $\pi^+_{\mbox{\tiny OFF}}=\$0.20$/kWh as on- and off-peak prices, respectively, and an {\em export} rate $\pi^-$ based on the wholesale market price.\footnote{Using the averaged 2018 real-time wholesale prices in Texas (\href{https://www.ercot.com/mktinfo/prices}{ERCOT}).} The DSO operating envelopes were assumed to be homogeneous with $\overline{z}_i = -\underline{z}_i = 6kW, \forall i\in \mathcal{N}$.

For every $i \in \mathcal{N}$, the household's consumption preferences are characterized by a quadratic concave and non-decreasing utility function of the form, for every device $k \in \mathcal{K}$,
\begin{equation}\label{eq:UtilityForm}
   U_{ik}(d_{ik})=\left\{\begin{array}{ll}
\alpha_{ik} d_{ik}-\frac{1}{2}\beta_{ik} d_{ik}^2,\hspace{-0.2cm} &\hspace{-0.2cm} 0 \leq d_{ik} \leq \frac{\alpha_{ik}}{\beta_{ik}} \\
\frac{\alpha_{ik}^2}{2 \beta_{ik}},\hspace{-0.2cm} &\hspace{-0.2cm} d_{ik}>\frac{\alpha_{ik}}{\beta_{ik}},
\end{array} \right.
\end{equation}
where $\alpha_{ik}, \beta_{ik}$ are parameters that are learned and calibrated using historical retail prices,\footnote{We used \href{https://data.austintexas.gov/stories/s/EOA-C-5-a-Austin-Energy-average-annual-system-rate/t4es-hvsj/}{Data.AustinTexas.gov} historical residential rates in Austin, TX.} consumption,\footnote{We used pre-2018 PecanStreet data for households in Austin, TX.} and elasticities for each load type (see appendix D in \cite{Alahmed&Tong:22EIRACM}).

Figure \ref{fig:NumWel} shows the average daily welfare gain (left) and normalized difference (right) of the centralized (\ref{eq:CentralOpt}) and decentralized (\ref{eq:DecentralOpt}) scheduling algorithms over the aggregate maximum welfare of standalone DSO customers, as the coalition size is increased by randomly adding households from the 20 households data. For every coalition size, 1000 Monte Carlo runs are implemented over the randomly selected households.

Two main observations are in order. First, as the coalition size increased, under both scheduling algorithms, the welfare difference and gain increased, which is due to the superadditivity of the value function under both schemes (Lemmas \ref{lem:AdditiveCentral}--\ref{lem:AdditiveDecentral}). The gain, however, saturated as coalition size increased, which was at 2.45\% and 1.55\% for centralized and decentralized scheduling, respectively. Also, as expected, the welfare difference, hence gain, of the centralized scheduling dominates decentralized scheduling. Second, we observe that although DER was decentrally scheduled under D-NEM, the community achieved welfare gains equivalent to centralized scheduling. Moreover, re-distributing the welfare of larger coalitions in Fig.\ref{fig:NumWel} is more computationally intensive under well-established {\em ex-post} allocations such as Shapley and the nucleolus; an issue that is mitigated under D-NEM.

\begin{figure}
    \centering
    \includegraphics[scale=0.38]{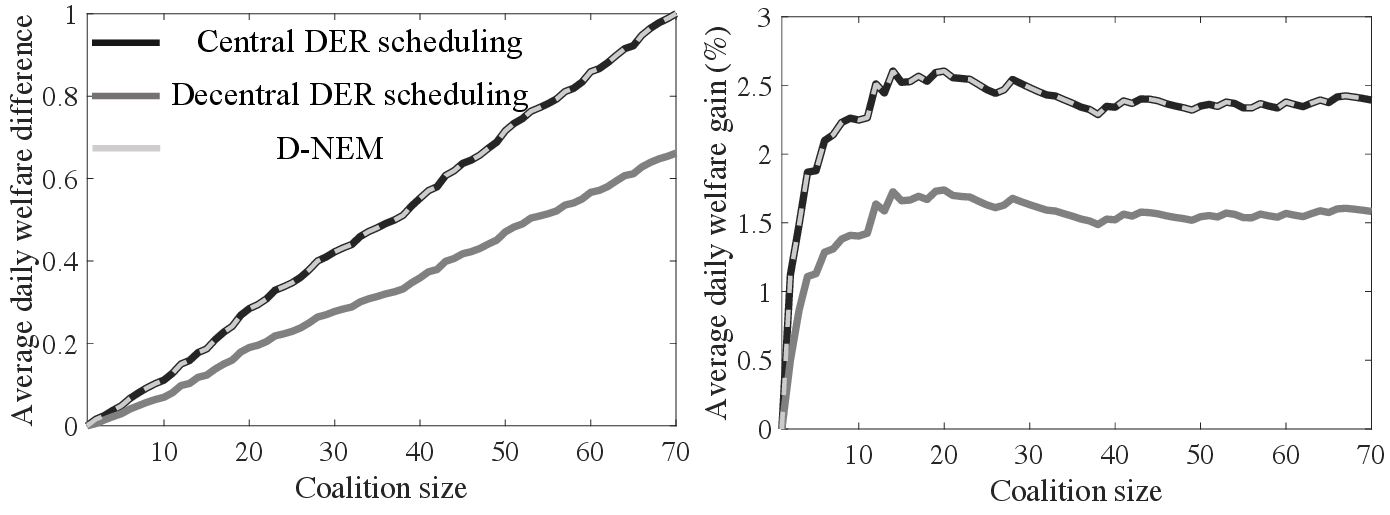}
    \setlength{\abovecaptionskip}{-0.5cm}
    \caption{Average daily welfare normalized difference and gain (\%).}
    \label{fig:NumWel}
    \vspace{-0.4cm}
\end{figure}

Table \ref{tab:NumIndRat} delves into the payoff allocation of the five presented schemes in Sec.\ref{sec:Allocation} and D-NEM under both centrally and decentrally scheduled coalitions, and presents how often the individual rationality axiom was violated, \ie number of hours whereby individual rationality fails $x$ over number of total hours, hence $x/(24\times 365\times |\Nc|)$. Under decentralized scheduling, only equal division and proportional allocations failed in achieving individual rationality, which necessarily mean that they do not belong to the core of the game. 

Under centralized scheduling, however, all of the {\em ex-post} allocation schemes failed in achieving individual rationality, meaning that they are not stabilizing allocations. Only the {\em ex-ante} D-NEM achieved individual rationality. We also observe that the frequency of violating individual rationality was consistently higher under the centralized scheduling over the decentralized one, which is because the community payment is subadditive only under the decentralized scheduling. Lastly, Table \ref{tab:NumIndRat} shows that the frequency of violating individual rationality by the {\em ex-post} allocations increased when $|\Nc|$ increased.

\begin{table}[]
\centering
\caption{Percentage of time of violating individual rationality (\%).}
\label{tab:NumIndRat}
\resizebox{\columnwidth}{!}{%
\begin{tabular}{@{}cccccccc@{}}
\toprule \midrule
$|\Nc|$        & DER Schedule  & Eq. division & Egalitarian & Proportional & \cite{Chakraborty&Poolla&Varaiya:19TSG} & Shapley & D-NEM \\ \midrule
\multirow{2}{*}{4}  & Centralized   & 55.3           & 5.0         & 27.8         & 2.5                  & 5.5     & 0     \\
                    & Decentralized & 55.1           & 0           & 27.6         & 0                    & 0       & 0     \\ \midrule
\multirow{2}{*}{10} & Centralized   &     61.1           &     5.1        &  30.1            &           3.0           &    5.6     & 0     \\
                    & Decentralized &         60.4       &       0      &     29.8         &     0                 &     0    & 0     \\ \midrule \bottomrule
\end{tabular}%
}
\vspace{-0.55cm}
\end{table}

\begin{figure}
    \centering
    \includegraphics[scale=0.25]{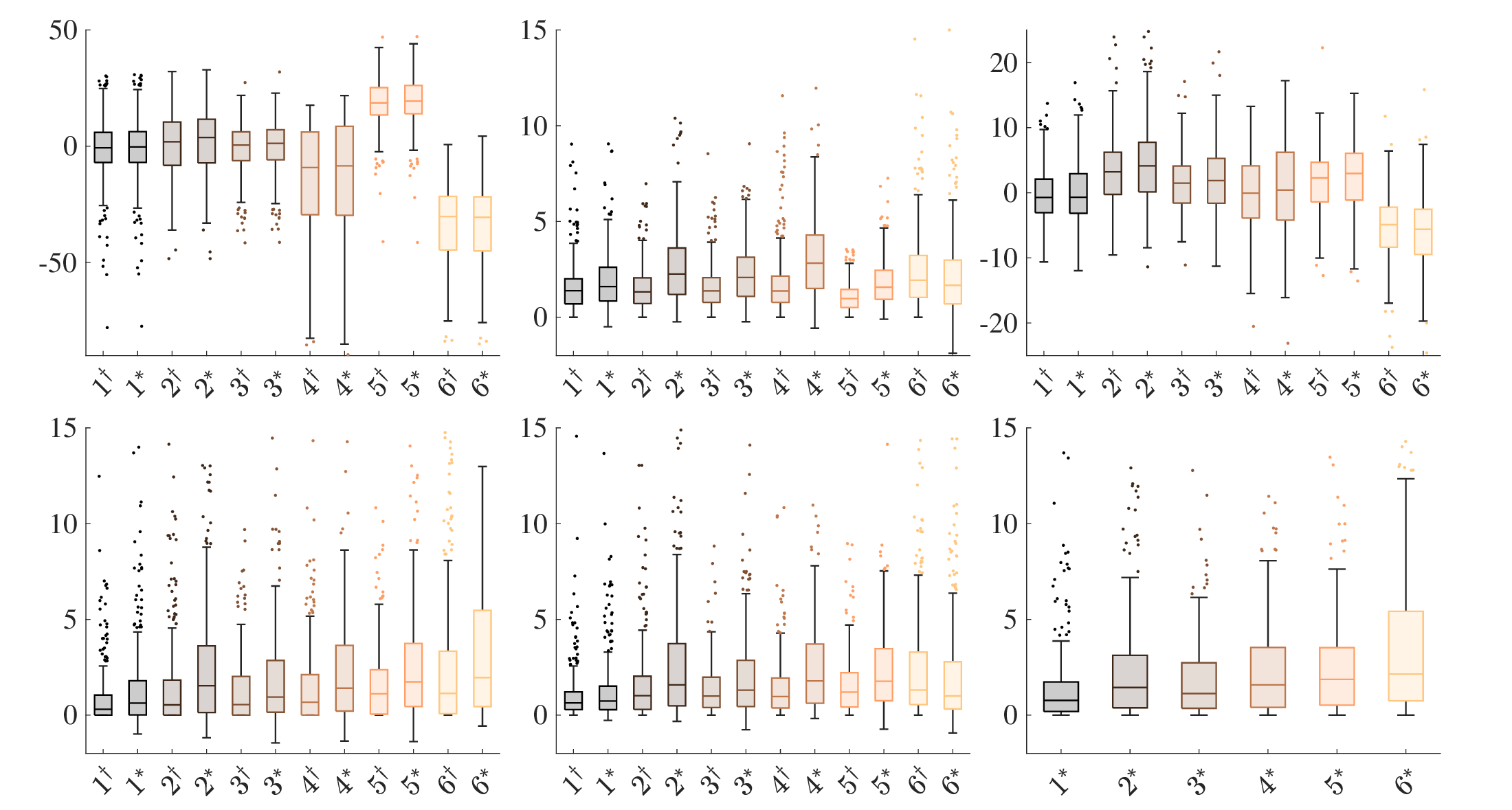}
    \setlength{\abovecaptionskip}{-0.5cm}
    \caption{Daily welfare gain (\%) of players ($|\Nc| = 6$) under decentralized$^\dagger$ and centralized$^\ast$ DER controls. Top panels (left to right): equal division, egalitarian, proportional. Bottom panels (left to right): \cite{Chakraborty&Poolla&Varaiya:19TSG}, Shapley, D-NEM.}
    \label{fig:NumBoxPlot}
    \vspace{-0.2cm}
\end{figure}

The daily welfare gain of each player in a coalition of size $|\Nc| = 6$ under both DER management algorithms and under the {\em ex-post} and {\em ex-ante} allocation mechanisms is shown in Fig.\ref{fig:NumBoxPlot}. One can see that the egalitarian, \cite{Chakraborty&Poolla&Varaiya:19TSG}, Shapley, and D-NEM allocations presented similar patterns of welfare distribution. D-NEM has only six boxplots because the gain is the same under both DER controls. Under both DER control algorithms, individual rationality was violated for all 6 customers under the equal division and proportional allocations. All 6 customers were individually rational under decentrally controlled coalitions and egalitarian, \cite{Chakraborty&Poolla&Varaiya:19TSG}, and Shapley allocations. These three allocations however fail to achieve individual rationality (particularly, customers 4--6) under the centralized DER control case, which maximizes the community welfare. The {\em ex-ante} D-NEM mechanism is the only policy that achieved individual rationality when the community welfare was maximized.

\section{Conclusion}\label{sec:conclusion}
 In this paper, a cooperative game theoretic approach is employed to analyze resource sharing and welfare distribution in energy community coalitions. We showed that when the community DER is centrally controlled to maximize the community welfare, and when the DER of each player are locally controlled to maximize each player's surplus, the cooperation value function is superadditive and the game is balanced. The payoff of the game is then allocated using five different {\em ex-post} payoff allocation mechanisms, which are compared to an {\em ex-ante} and welfare maximizing mechanism. In simulations, we evaluated the welfare gains of the energy coalition under both scheduling algorithms and the performance of the {\em ex-post} and {\em ex-ante} allocation mechanisms.
{
\bibliographystyle{IEEEtran}
\bibliography{ORIE7191}

\begin{thebibliography}{10}
\providecommand{\url}[1]{#1}
\csname url@samestyle\endcsname
\providecommand{\newblock}{\relax}
\providecommand{\bibinfo}[2]{#2}
\providecommand{\BIBentrySTDinterwordspacing}{\spaceskip=0pt\relax}
\providecommand{\BIBentryALTinterwordstretchfactor}{4}
\providecommand{\BIBentryALTinterwordspacing}{\spaceskip=\fontdimen2\font plus
\BIBentryALTinterwordstretchfactor\fontdimen3\font minus \fontdimen4\font\relax}
\providecommand{\BIBforeignlanguage}[2]{{%
\expandafter\ifx\csname l@#1\endcsname\relax
\typeout{** WARNING: IEEEtran.bst: No hyphenation pattern has been}%
\typeout{** loaded for the language `#1'. Using the pattern for}%
\typeout{** the default language instead.}%
\else
\language=\csname l@#1\endcsname
\fi
#2}}
\providecommand{\BIBdecl}{\relax}
\BIBdecl

\bibitem{Alahmed&Tong:22EIRACM}
\BIBentryALTinterwordspacing
A.~S. Alahmed and L.~Tong, ``Integrating distributed energy resources: Optimal prosumer decisions and impacts of net metering tariffs,'' \emph{SIGENERGY Energy Inform. Rev.}, vol.~2, no.~2, p. 13–31, Aug. 2022. [Online]. Available: \url{https://doi.org/10.1145/3555006.3555008}
\BIBentrySTDinterwordspacing

\bibitem{Alahmed&Tong:23ECjournalarXiv}
\BIBentryALTinterwordspacing
------, ``Dynamic net metering for energy communities,'' 2023. [Online]. Available: \url{https://arxiv.org/abs/2306.13677}
\BIBentrySTDinterwordspacing

\bibitem{Yang&Guoqiang&Spanos:21TSG}
Y.~Yang, G.~Hu, and C.~J. Spanos, ``Optimal sharing and fair cost allocation of community energy storage,'' \emph{IEEE Transactions on Smart Grid}, vol.~12, no.~5, pp. 4185--4194, Sep. 2021.

\bibitem{Chakraborty&Baeyens&Khargonekar:TPS18}
P.~Chakraborty, E.~Baeyens, and P.~P. Khargonekar, ``Cost causation based allocations of costs for market integration of renewable energy,'' \emph{IEEE Transactions on Power Systems}, vol.~33, no.~1, pp. 70--83, 2018.

\bibitem{Han&Morstyn&McCulloch:19TPS}
L.~Han, T.~Morstyn, and M.~McCulloch, ``Incentivizing prosumer coalitions with energy management using cooperative game theory,'' \emph{IEEE Transactions on Power Systems}, vol.~34, no.~1, pp. 303--313, 2019.

\bibitem{Fleischhacker&Corinaldesi&Lettner&Auer&Botterud:22TSG}
A.~Fleischhacker, C.~Corinaldesi, G.~Lettner, H.~Auer, and A.~Botterud, ``Stabilizing energy communities through energy pricing or {PV} expansion,'' \emph{IEEE Trans. on Smart Grid}, vol.~13, no.~1, pp. 728--737, 2022.

\bibitem{Cui&Wang&Yan&Shi&Xiao:21TSG}
S.~Cui, Y.-W. Wang, Y.~Shi, and J.-W. Xiao, ``Community energy cooperation with the presence of cheating behaviors,'' \emph{IEEE Transactions on Smart Grid}, vol.~12, no.~1, pp. 561--573, 2021.

\bibitem{Vespermann&Hamacher&Kazempour:21TPS}
N.~Vespermann, T.~Hamacher, and J.~Kazempour, ``{Access Economy for Storage in Energy Communities},'' \emph{IEEE Transactions on Power Systems}, vol.~36, no.~3, pp. 2234--2250, 2021.

\bibitem{Chakraborty&Poolla&Varaiya:19TSG}
P.~{Chakraborty}, E.~{Baeyens}, P.~P. {Khargonekar}, K.~{Poolla}, and P.~{Varaiya}, ``Analysis of solar energy aggregation under various billing mechanisms,'' \emph{IEEE Tran. on Smart Grid}, vol.~10, no.~4, pp. 4175--4187, 2019.

\bibitem{Alahmed&Cavraro&Bernstein&Tong:23AllertonArXiv}
A.~S. Alahmed, G.~Cavraro, A.~Bernstein, and L.~Tong, ``Operating-envelopes-aware decentralized welfare maximization for energy communities,'' in \emph{2023 59th Annual Allerton Conference on Communication, Control, and Computing (Allerton)}, 2023, pp. 1--8.

\bibitem{Alahmed&Tong:22IEEETSG}
A.~S. Alahmed and L.~Tong, ``On net energy metering {X}: Optimal prosumer decisions, social welfare, and cross-subsidies,'' \emph{IEEE Transactions on Smart Grid}, vol.~14, no.~02, pp. 1652--1663, 2023.

\bibitem{Shapley:63NRLQ}
\BIBentryALTinterwordspacing
L.~S. Shapley, ``On balanced sets and cores,'' \emph{Naval Research Logistics Quarterly}, vol.~14, no.~4, pp. 453--460, 1967. [Online]. Available: \url{https://onlinelibrary.wiley.com/doi/abs/10.1002/nav.3800140404}
\BIBentrySTDinterwordspacing

\bibitem{HPYoung:85CostAllocationBook}
\BIBentryALTinterwordspacing
H.~P. Young, \emph{Cost Allocation: Methods, Principles, Applications}.\hskip 1em plus 0.5em minus 0.4em\relax Amsterdam: North Holland Publishing Co., March 1985. [Online]. Available: \url{https://pure.iiasa.ac.at/id/eprint/2596/}
\BIBentrySTDinterwordspacing

\end{thebibliography}
}
\section*{Appendix: Proofs}\label{sec:appendix_axioms}
\subsection{Proof of Lemma \ref{lem:AdditiveCentral}}\label{subsec:lemAdditiveCentral}
We need to show that for all $\Nc,\Tc\subseteq \Hc,~ \Nc \cap \Tc=\emptyset$,
\begin{align}
    \nu^\ast(\Nc)+\nu^\ast(\Tc) &\leq \nu^\ast(\Nc \cup \Tc).
\end{align}
Note that
\begin{align}
     \nu^\ast(\Nc) &= W^\ast(\bm{d}^{\Nc^\ast}, \bm{z}^{\Nc^\ast}) = \sum_{i\in \Nc} U_i(\bm{d}^{\Nc^\ast}_i) - P^\pi(\sum_{i\in \Nc}z^{\Nc^\ast}_i)\label{eq:lem1_1}\\
     \nu^\ast(\Tc) &= W^\ast(\bm{d}^{\Tc^\ast}, \bm{z}^{\Tc^\ast}) = \sum_{i\in \Tc} U_i(\bm{d}^{\Tc^\ast}_i) - P^\pi(\sum_{i\in \Tc}z^{\Tc^\ast}_i),\label{eq:lem1_2}
\end{align}
where for any coalition $\Mc \in 2^{\Hc}, \bm{d}^{\Mc^\ast}:=\{\bm{d}^{\Mc^\ast}_i\}, \bm{z}^{\Mc^\ast}:=\{z^{\Mc^\ast}_i\}, \forall i\in \Mc$, and the second term in (\ref{eq:lem1_1})-(\ref{eq:lem1_2}) above is from leveraging constraint (\ref{eq:ProfitNeut}). If coalitions $\Nc$ and $\Tc$ form one coalition $\Nc \cup \Tc$ and continue scheduling, as before, their new coalition's value becomes 
\begin{align}
    &\nu({\Nc} \cup {\Tc}) = W(\bm{d}^{\Nc^\ast \cup \Tc^\ast}, \bm{z}^{\Nc^\ast \cup \Tc^\ast}) \nn\\&=  \sum_{i\in \Nc \cup \Tc} U_i(\bm{d}^{\Nc^\ast \cup \Tc^\ast}_i) - P^\pi(\sum_{i\in \Nc \cup \Tc} z^{\Nc^\ast \cup \Tc^\ast}_i)\nn\\&= \sum_{i\in \Nc} U_i(\bm{d}^{\Nc^\ast}_i) + \sum_{i\in \Tc} U_i(\bm{d}^{\Tc^\ast}_i)- P^\pi(\sum_{i\in \Nc}z^{\Nc^\ast}_i+\sum_{i\in \Tc}z^{\Tc^\ast}_i).\nn
\end{align}
Note that because
\begin{align*}
    &\nu({\Nc} \cup {\Tc}) - \nu^\ast(\Nc) - \nu^\ast(\Tc) \nn\\&=   P^\pi(\sum_{i\in \Nc}z^{\Nc^\ast}_i) + P^\pi(\sum_{i\in \Tc}z^{\Tc^\ast}_i)  - P^\pi(\sum_{i\in \Nc}z^{\Nc^\ast}_i+\sum_{i\in \Tc}z^{\Tc^\ast}_i)\nn\\&\stackrel{\text{(Lem.\ref{lem:AdditiveDecentral})}}{\geq} 0 ,
\end{align*}
we have 
\begin{equation}\label{eq:lem1_3}
    \nu({\Nc} \cup {\Tc}) \geq \nu^\ast(\Nc) + \nu^\ast(\Tc) .
\end{equation}
Now, by solving (\ref{eq:CentralOpt}) for coalition $\Nc \cup \Tc$ we have
\begin{align*}
\nu^\ast(\Nc \cup \Tc) &=  W^\ast(\bm{d}^{{(\Nc \cup \Tc)}^\ast},\bm{z}^{{(\Nc \cup \Tc)}^\ast}) \nn\\&\geq W(\bm{d}^{\Nc^\ast \cup \Tc^\ast}, \bm{z}^{\Nc^\ast \cup \Tc^\ast}) = \nu({\Nc} \cup {\Tc})\nn\\&\stackrel{\text{(\ref{eq:lem1_3})}}{\geq}
\nu^\ast(\Nc)+\nu^\ast(\Tc),
\end{align*}
which completes the proof. \hfill$\blacksquare$

\subsection{Proof of Lemma \ref{lem:AdditiveDecentral}}\label{subsec:lemAdditiveDecentral}
Given that the players' face the same NEM X tariff design regardless of the community size, their optimal consumption and net consumption schedules ($\bm{d}^\dagger_i, z^\dagger_i$) stay the same. Therefore, for any $\Nc,\Tc \subseteq \Hc$, we have 
\begin{align}
\nu^\dagger(\Nc \cup \Tc) &= W^\dagger(\Nc \cup \Tc)= \sum_{i\in \Nc \cup \Tc} U_i(\bm{d}^\dagger_i) - P^\pi(z^\dagger_{\Nc}+z^\dagger_{\Tc}).\nn\\
\nu^\dagger(\Nc) &= W^\dagger(\Nc) = \sum_{i\in \Nc} U_i(\bm{d}^\dagger_i) - P^\pi(z^\dagger_{\Nc})\nn\\
\nu^\dagger(\Tc) &= W^\dagger(\Tc) = \sum_{i\in \Tc} U_i(\bm{d}^\dagger_i) - P^\pi(z^\dagger_{\Tc}),\nn
\end{align}
where for any coalition $\Mc \in 2^{\Hc}$, $z^\dagger_{\Mc}:= \sum_{i\in \Mc} z_i^\dagger$. Given that the utility function terms cancel out, the proof is complete if we show the following
\begin{align*}
    \nu^\dagger(\Nc)+\nu^\dagger(\Tc) &\leq \nu^\dagger(\Nc \cup \Tc),
\end{align*}
which can reformulated to
\begin{align}
    P^\pi(z^\dagger_{\Nc \cup \Tc}) \leq P^\pi(z^\dagger_{\Nc})+ P^\pi(z^\dagger_{\Tc}).\label{eq:lemdecentralineq}
\end{align}
where $z^\dagger_{\Nc \cup \Tc} := z^\dagger_{\Nc}+z^\dagger_{\Tc}$.
We prove (\ref{eq:lemdecentralineq}) using the following four exhaustive cases.
\begin{itemize}[wide, labelwidth=!, labelindent=0pt]
    \item Case 1 ($z^\dagger_\Nc \geq 0, z^\dagger_\Tc \geq 0$): This leads to $z^\dagger_{\Nc \cup \Tc} \geq 0$. The LHS and RHS in (\ref{eq:lemdecentralineq}) are, respectively:
    \begin{align*}
        P^\pi(z^\dagger_{\Nc \cup \Tc}) &= \pi^+ (z^\dagger_\Nc + z^\dagger_\Tc)\\
    P^\pi(z^\dagger_{\Nc})+ P^\pi(z^\dagger_{\Tc})&=\pi^+ z^\dagger_\Nc + \pi^+ z^\dagger_\Tc = \pi^+ (z^\dagger_\Nc + z^\dagger_\Tc),
    \end{align*}
    hence (\ref{eq:lemdecentralineq}) holds.
    \item Case 2 ($z^\dagger_\Nc \leq 0, z^\dagger_\Tc \leq 0$): This leads to $z^\dagger_{\Nc \cup \Tc} \leq 0$. The LHS and RHS in (\ref{eq:lemdecentralineq}) are, respectively:
    \begin{align*}
        P^\pi(z^\dagger_{\Nc \cup \Tc}) &= \pi^- (z^\dagger_\Nc + z^\dagger_\Tc)\\
    P^\pi(z^\dagger_{\Nc})+ P^\pi(z^\dagger_{\Tc})&=\pi^- z^\dagger_\Nc + \pi^- z^\dagger_\Tc = \pi^- (z^\dagger_\Nc + z^\dagger_\Tc),
    \end{align*}
    hence (\ref{eq:lemdecentralineq}) holds.
    \item Case 3 ($z^\dagger_\Nc \geq 0, z^\dagger_\Tc \leq 0$): We have two sub-cases here
    \begin{itemize}
    \item Case 3.1 ($z^\dagger_{\Nc \cup \Tc} \geq 0$): The LHS and RHS in (\ref{eq:lemdecentralineq}) are, respectively:
    \begin{align*}
        P^\pi(z^\dagger_{\Nc \cup \Tc}) &= \pi^+ (z^\dagger_\Nc + z^\dagger_\Tc)\\
    P^\pi(z^\dagger_{\Nc})+ P^\pi(z^\dagger_{\Tc})&= \pi^+ z^\dagger_\Nc + \pi^- z^\dagger_\Tc,
    \end{align*}
By computing the difference 
\begin{align*}
    P^\pi(z^\dagger_{\Nc \cup \Tc}) - P^\pi(z^\dagger_{\Nc})- P^\pi(z^\dagger_{\Tc}) = (\pi^+-\pi^-) z^\dagger_\Tc \leq 0,
\end{align*}
we verify that (\ref{eq:lemdecentralineq}) holds, as $\pi^-\leq \pi^+$.
    \item Case 3.2 ($z^\dagger_{\Nc \cup \Tc} \leq 0$): Following the same steps in Case 3.1, we get
    \begin{align*}
    P^\pi(z^\dagger_{\Nc \cup \Tc}) - P^\pi(z^\dagger_{\Nc})- P^\pi(z^\dagger_{\Tc}) = (\pi^--\pi^+) z^\dagger_\Nc \leq 0,
\end{align*}
which verifies that (\ref{eq:lemdecentralineq}) holds, as $\pi^-\leq \pi^+$.
    \end{itemize}
    \item Case 4 ($z^\dagger_\Nc \leq 0, z^\dagger_\Tc \geq 0$): We have two sub-cases here
    \begin{itemize}
    \item Case 4.1 ($z^\dagger_{\Nc \cup \Tc} \geq 0$): The LHS and RHS in (\ref{eq:lemdecentralineq}) are, respectively:
    \begin{align*}
        P^\pi(z^\dagger_{\Nc \cup \Tc}) &= \pi^+ (z^\dagger_\Nc + z^\dagger_\Tc)\\
    P^\pi(z^\dagger_{\Nc})+ P^\pi(z^\dagger_{\Tc})&= \pi^- (z^\dagger_\Nc) + \pi^+ (z^\dagger_\Tc),
    \end{align*}
By computing the difference 
\begin{align*}
    P^\pi(z^\dagger_{\Nc \cup \Tc}) - P^\pi(z^\dagger_{\Nc})- P^\pi(z^\dagger_{\Tc}) = (\pi^+-\pi^-) z^\dagger_\Nc \leq 0,
\end{align*}
we verify that (\ref{eq:lemdecentralineq}) holds, as $\pi^-\leq \pi^+$.
    \item Case 4.2 ($z^\dagger_{\Nc \cup \Tc} \leq 0$): Following the same steps in Case 4.1, we get
    \begin{align*}
    P^\pi(z^\dagger_{\Nc \cup \Tc}) - P^\pi(z^\dagger_{\Nc})- P^\pi(z^\dagger_{\Tc}) = (\pi^--\pi^+) z^\dagger_\Tc \leq 0,
\end{align*}
which verifies that (\ref{eq:lemdecentralineq}) holds, as $\pi^-\leq \pi^+$.
    \end{itemize}
\end{itemize}
The four cases prove (\ref{eq:lemdecentralineq}), which says that the payment function under the decentralized scheduling ($\bm{d}^\dagger_i, z^\dagger_i$) is subadditive, hence $
    \nu^\dagger(\Nc)+\nu^\dagger(\Tc) \leq \nu^\dagger(\Nc \cup \Tc).$
\hfill$\blacksquare$

\begin{lemma}\label{lem:homogenity}
    For any coalition $\Nc \in 2^{\Hc}$, the value function under centralized DER scheduling $ \nu^\ast(\Nc)= W^\ast(\bm{d}^{\Nc^\ast}, \bm{z}^{\Nc^\ast})$ obeys
    \begin{equation}
       W^\ast(\beta \bm{d}^{\Nc^\ast}, \beta\bm{z}^{\Nc^\ast}) \geq \beta W^\ast(\bm{d}^{\Nc^\ast}, \bm{z}^{\Nc^\ast}),
    \end{equation}
    where $\beta \in[0,1]$.
\end{lemma}

\subsection{Lemma \ref{lem:homogenity} and Proof of Lemma \ref{lem:homogenity}}
Recall the welfare maximization problem under centralized DER scheduling
\begin{subequations} 
\begin{align*} 
W^\ast(\bm{d}^{\Nc^\ast},\bm{z}^{\Nc^\ast}):=  \underset{\{\bm{d}_{i}\}_{i=1}^{|\Nc|},\{z_{i}\}_{i=1}^{|\Nc|}}{\text { maximize}} &~~ W(\Nc):=\sum_{i\in \Nc} S_{i}^{\pi}(\bm{d}_{i},z_{i}) \\\text { subject to~~} & \sum_{i\in \Nc} P^\pi(z_{i}) = P^\pi(\sum_{i\in \Nc} z_{i}) 
\\& \text{for all }~ i=1,\ldots, N\nn
\\& z_i = \bm{1}^\top \bm{d}_{i} - r_{i} 
\\&  \underline{z}_i \leq z_{i} \leq \overline{z}_i
\\& \underline{\bm{d}}_i \preceq \bm{d}_{i} \preceq \overline{\bm{d}}_i,
\end{align*} \vspace{-0.1cm}
\end{subequations}
which can be more compactly written as
\begin{align*}
    W^\ast(\bm{d}^{\Nc^\ast},\bm{z}^{\Nc^\ast}):= \underset{(\bm{d},\bm{z}) \in \Omega^{\Nc}}{\text { maximize}} \sum_{i\in \Nc} U_i(\bm{d}_i) - P^\pi(\sum_{i\in \Nc} z_{i}),
\end{align*}
where we used the constraint (\ref{eq:ProfitNeut}), the notation $\bm{d}^{\Mc^\ast}:=\{\bm{d}^{\Mc^\ast}_i\}, \bm{z}^{\Mc^\ast}:=\{z^{\Mc^\ast}_i\}, \forall i\in \Mc \in 2^{\Hc}$, and the feasible set notation 
\begin{align*}
    \Omega^{\Nc}:=\{&(\bm{d}_1,\ldots,\bm{d}_N),(z_1,\ldots,z_N) : z_i = \bm{1}^\top \bm{d}_{i} - r_{i},\\& \underline{z}_i \leq z_{i} \leq \overline{z}_i, \underline{\bm{d}}_i \preceq \bm{d}_{i} \preceq \overline{\bm{d}}_i \text{ for all } i=1,\ldots,N\}.
\end{align*}
Given that the constraints are linear/affine, we can scale the decision variables by $\beta$ without changing the feasible set $\Omega^{\Nc}$.

Now, we solve
\begin{align*}
    W^\ast(\beta \bm{d}^{\Nc^\ast}\hspace{-0.1cm},\beta \bm{z}&^{\Nc^\ast})\hspace{-0.1cm}=\hspace{-0.1cm} \underset{(\bm{d},\bm{z}) \in \Omega^{\Nc}}{\text {maximize}} \bigg( \sum_{i\in \Nc} U_i(\beta \bm{d}_i) - P^\pi( \sum_{i\in \Nc} \beta z_{i})\bigg)\\&\stackrel{\text{(A)}}{=} \underset{(\bm{d},\bm{z}) \in \Omega^{\Nc}}{\text {maximize}} \bigg(\sum_{i\in \Nc} U_i(\beta \bm{d}_i) - \beta P^\pi( \sum_{i\in \Nc} z_{i})\bigg) \\&\stackrel{\text{(B)}}{\geq} \underset{(\bm{d},\bm{z}) \in \Omega^{\Nc}}{\text {maximize}} \bigg(\beta \sum_{i\in \Nc} U_i( \bm{d}_i) - \beta P^\pi( \sum_{i\in \Nc} z_{i})\bigg) \\&=\beta\cdot \underset{(\bm{d},\bm{z}) \in \Omega^{\Nc}}{\text {maximize}}  \bigg(\sum_{i\in \Nc} U_i( \bm{d}_i) - P^\pi(\sum_{i\in \Nc} z_{i}) \bigg)\\&= \beta W^\ast(\bm{d}^{\Nc^\ast},\bm{z}^{\Nc^\ast}),
\end{align*}
where step (A) is because, for any $x\in \mathbb{R}$ and $\beta \in[0,1]$,
\begin{align*}
P^\pi(\beta x) &= \pi^+ [\beta x]^+ -\pi^- [\beta x]^-\\&= \beta \pi^+[x]^+-\beta \pi^-[x]^- = \beta P^\pi(x),
\end{align*}
and step (B) is due to the concavity, additivity and $U_i(\bm{0})=0$ properties of $U_i(\cdot)$, which yield, for all $i \in \Nc, k\in \Kc$,
\begin{align*}
    U_{ik}(\beta d_{ik}) &= U_{ik}(0+\beta d_{ik})\\& \geq (1-\beta)U_{ik}(0)+\beta U_{ik}(d_{ik}) =\beta U_{ik}(d_{ik}),
\end{align*}
which completes the proof. \hfill$\blacksquare$

\subsection{Proof of Theorem \ref{thm:BalancedCentral}}\label{subsec:thmAdditiveCentral}
We prove Theorem \ref{thm:BalancedCentral} by leveraging Lemma \ref{lem:AdditiveCentral} and Lemma \ref{lem:homogenity}. The balancedness of the cooperative game under centralized DER scheduling ($\Hc,\nu^\ast$) is proved, if we show
\begin{equation}\label{eq:Thm1Def}
    \sum_{\Nc \in 2^\Hc} \alpha(\Nc) \nu^\ast(\Nc) \leq \nu^\ast(\Hc).
\end{equation}
By reformulating the LHS of (\ref{eq:Thm1Def}), we get
\begin{align*}
    &\sum_{\Nc \in 2^\Hc} \alpha(\Nc) \nu^\ast(\Nc)= \sum_{\Nc \in 2^\Hc} \alpha(\Nc) W^\ast(\bm{d}^{\Nc^\ast}, \bm{z}^{\Nc^\ast})\\&\stackrel{\text{(Lem.\ref{lem:homogenity})}}{\leq} \sum_{\Nc \in 2^\Hc}  W^\ast(\alpha(\Nc) \bm{d}^{\Nc^\ast}, \alpha(\Nc) \bm{z}^{\Nc^\ast})\\&\stackrel{\text{(Lem.\ref{lem:AdditiveCentral})}}{\leq}
    W^\ast(\sum_{\Nc \in 2^\Hc} \alpha(\Nc) \bm{d}^{\Nc^\ast}, \sum_{\Nc \in 2^\Hc} \alpha(\Nc) \bm{z}^{\Nc^\ast})\\&=
    W^\ast(\sum_{\Nc \in 2^{\Hc}} \alpha(\Nc) \bm{1}_{\Nc}(i) \bm{d}^{\Hc^\ast}, \sum_{\Nc \in 2^{\Hc}} \alpha(\Nc)  \bm{1}_{\Nc}(i) \bm{z}^{\Hc^\ast}) 
    \\&\stackrel{\text{(Def.\ref{def:BalancedGames})}}{=} W^\ast(\bm{d}^{\Hc^\ast}, \bm{z}^{\Hc^\ast}) = \nu^\ast(\Hc),
\end{align*}
which completes the proof. \hfill$\blacksquare$

\subsection{Proof of Theorem \ref{thm:BalancedDecentral}}\label{subsec:thmAdditiveDecentral}
We prove the balancedness of the game ($\Hc,\nu^\dagger$) using the subadditivity of the payment function under decentralized scheduling ($\bm{d}^\dagger_i, z^\dagger_i$) shown in Lemma \ref{lem:AdditiveDecentral}. From Def.\ref{def:BalancedGames} we need to show that
\begin{equation}\label{eq:Thm2Def}
    \sum_{\Nc \in 2^\Hc} \alpha(\Nc) \nu^\dagger(\Nc) \leq \nu^\dagger(\Hc).
\end{equation}
By reformulating the LHS of (\ref{eq:Thm2Def}), we get
\begin{align*}
    &\sum_{\Nc \in 2^\Hc} \alpha(\Nc) \nu^\dagger(\Nc)\\ &=  \sum_{\Nc \in 2^\Hc} \alpha(\Nc) W^\dagger(\Nc) \stackrel{\text{(\ref{eq:Welfare1Mod})}}{=} \sum_{\Nc \in 2^\Hc} \alpha(\Nc) \bigg(\sum_{i\in \Nc} U_i(\bm{d}_i^\dagger ) - P^\pi (z_{\Nc}^\dagger)\bigg)\\&= 
   \sum_{i\in \Hc} \sum_{\Nc \in 2^\Hc} \alpha(\Nc) \bm{1}_{\Nc}(i) U_i(\bm{d}_i^\dagger) - \sum_{\Nc \in 2^\Hc} \alpha(\Nc) P^\pi (z_{\Nc}^\dagger)\\& \stackrel{\text{(Def.\ref{def:BalancedGames})}}{=} \sum_{i\in \Hc} U_i(\bm{d}_i^\dagger ) - \sum_{\Nc \in 2^\Hc} \alpha(\Nc) P^\pi (z_{\Nc}^\dagger)\\&\stackrel{\text{(A)}}{=} \sum_{i\in \Hc} U_i(\bm{d}_i^\dagger ) - \sum_{\Nc \in 2^\Hc} P^\pi (\alpha(\Nc) z_{\Nc}^\dagger)\\& \stackrel{\text{(Lem.\ref{lem:AdditiveDecentral})}}{\leq} \sum_{i\in \Hc} U_i(\bm{d}_i^\dagger ) - P^\pi (\sum_{\Nc \in 2^\Hc} \alpha(\Nc) z_{\Nc}^\dagger)\\&= \sum_{i\in \Hc} U_i(\bm{d}_i^\dagger ) - P^\pi (\sum_{i\in \Hc} \sum_{\Nc \in 2^\Hc} \alpha(\Nc) \bm{1}_{\Nc}(i) z_{i}^\dagger)\\&\stackrel{\text{(Def.\ref{def:BalancedGames})}}{=}  \sum_{i\in \Hc} U_i(\bm{d}_i^\dagger ) - P^\pi(z_{\Hc}^\dagger) \stackrel{\text{(\ref{eq:Welfare1Mod})}}{=} W^\dagger(\Hc) = \nu^\dagger(\Hc),
\end{align*}
where in step (A) we used the homogeneity of $P^\pi(\cdot)$ under decentralized DER scheduling, \ie for $\Nc \in 2^{\Hc}$ and scalar $\beta \geq 0$,
$$P^\pi(\beta z_{\Nc}^\dagger) = \pi^+ [\beta z_\Nc]^+ - [\beta z_\Nc]^- = \beta P^\pi(z_{\Nc}^\dagger),$$
which completes the proof. \hfill$\blacksquare$
\end{document}